\documentclass[11pt]{article}
\usepackage{amsmath,amssymb,color,graphics,epsfig,cite}

\usepackage{amsfonts}

\usepackage{bm}
\usepackage{slashed}
\usepackage{framed}
\usepackage{physics}
\usepackage{hyperref}
\newcommand{\be}{\begin{equation}}
\newcommand{\ee}{\end{equation}}
\newcommand{\bea}{\setlength\arraycolsep{2pt} \begin{eqnarray}}
\newcommand{\eea}{\end{eqnarray}}

\def\0{{\sst{(0)}}}
\def\1{{\sst{(1)}}}
\def\2{{\sst{(2)}}}
\def\3{{\sst{(3)}}}
\def\4{{\sst{(4)}}}
\def\5{{\sst{(5)}}}
\def\6{{\sst{(6)}}}
\def\7{{\sst{(7)}}}
\def\8{{\sst{(8)}}}
\def\sst#1{{\scriptscriptstyle #1}}

\begin{document}

\begin{center}
{\Large {\bf AdS Wormholes from Ricci-flat/AdS Correspondence}
}

\vspace{20pt}

{\large Tianhao Wu}

\vspace{10pt}

{\it Department of Physics, University of Illinois, Urbana-Champaign,\\
 Urbana, 61801 United States\\}
{\it Center for Joint Quantum Studies and Department of Physics,\\
School of Science, Tianjin University, Tianjin 300350, China}

\vspace{40pt}

\underline{ABSTRACT}
\end{center}

We discuss the Wormholes in general dimensions by studying the Einstein-phantom scalar field with and without the cosmological constant. Solving AdS wormholes in general dimension is hard due to the nonlinear nature of the theory. In this work, we implement the AdS/Ricci-flat correspondence, extended to include the axion field(the phantom scalar field), to construct AdS wormholes. Wormholes of Ellis-Bronnikov class are discussed in general dimensions.

\vfill {\footnotesize  twu49@illinois.edu}


\thispagestyle{empty}
\pagebreak

\tableofcontents

\section{Introduction}
\label{sec:intro}

Wormhole is thought to be the tool for fast interstellar travel, for it connects different parts of space-times via the throat.\cite{Thorne, visual} Traversable wormholes have been studied in 4-dimension and higher dimensions.\cite{visser} The most intuitive wormhole is the Ellis wormhole from the theory $\mathcal{S} = \frac{1}{16\pi G} \int \dd x^4 \sqrt{-g} \big( R + \frac{1}{2}(\partial \chi)^2 \big)$. The singularity in Schiwartzchild space-time hinges the exploration of particle theories in gravitational fields as it is not geodesically complete. Ellis wormhole was constructed in the effort to remove the problematic singularities in Schwartzchild space-times.\cite{bronnikov, ellis, Einstein} It was first discussed in 4-dimension where it was illustrated as a drainhole. Ellis wormhole in higher dimensions was later studied in different settings.\cite{torii, LuHuang, review} Having the general ansatz in higher dimension, the general Ellis-Bronnikov class wormhole was also constructed.\cite{nozawa I, uniqueness static, uniqueness} Specific solutions of wormholes with bare cosmological constant in 4-dimension were found under various conditions.\cite{Morris4d, AdS4d} 

However, wormholes with bare cosmological constant in higher dimensions are yet to be constructed.\cite{nozawa III} Such AdS wormholes play crucial role since AdS/CFT correspondence becomes a rich field of study.\cite{juanAdS, Juan, Witten, LuPope} In the context of AdS/CFT correspondence, the dimension of conformal field theory on the boundary is of co-dimension 1 to the AdS gravity in the bulk with generic dimensions. Thus looking for AdS wormhole solutions in general dimension is of particular interest. In this paper, we use the Ricci-flat/AdS correspondence to construct the solutions of wormholes with bare cosmological constant in general dimensions from Ellis-Bronnikov wormhole solutions.

The AdS/Ricci-flat correspondence relates the solutions in asymptotically AdS space on a torus and asymptotically flat space on a sphere.\cite{KKskenderis, KKskenderis2} Kuluza-Klein reduction was used to demonstrate the validity of this correspondence. KK reduction comes from the string compactification. It reduces higher dimensional theories to lower dimensional ones, keeping the additional dimensions compactified.\cite{KKSugra} 
Previous works show that this correspondence was used to construct new solutions in supergravity theories,\cite{LuMa} where matter field was added to the AdS theories on torus. The correspondence interchanges the matter field with the Ricci scalar of the sphere in the pure gravity theory.

In this paper, we consider the action for the AdS wormhole is given by 
\begin{equation}
    \mathcal{S} = \frac{1}{16 \pi G} \int \dd x^n \sqrt{-g} \Big( R - 2 \Lambda + \frac{1}{2} ( \partial \chi )^2 \Big)
\end{equation} where $\chi$ is the phantom scalar field. In this paper we add matter fields, the phantom field $\chi$, to both theories in the correspondence. We found that the matter fields descend down after the KK reduction calculation for both theories. So the AdS/Ricci-flat correspondence interchanges the phantom field in Ellis-Bronnikov wormhole with matter field in AdS wormhole.
Such AdS wormhole needs to violate the null energy condition.\cite{NullEnergy, energy condition} The null energy condition is believed to be satisfied in general relativity. 
However, the Ellis wormhole is opened via the addition of ghost field, or what we call axion. So the criterion for a solution to be a wormhole is that the solution violates the null energy condition.  The solution constructed from AdS/Ricci-flat correspondence does satisfy this requirement.

The organisation of this paper is as follows. In section 2, we describe the correspondence of two theories and discuss the identification of parameters. In section 3, we find the solutions of wormholes with axion field, namely the Ellis wormhole, then we use the correspondence map to construct the solutions of AdS Ellis wormholes. In section 4, we further the discussion in last section to include generic solution of Ellis-Bronnikov wormhole. We see that in 4-dimension, this generic solution reduces back to the form in previous sections. In section 5, we find the solution of AdS wormhole in general dimension by applying AdS/Ricci-flat correspondence. We show this solution reduces back to the AdS Ellis wormhole via coordinate transformations. In section 6, We demonstrated that the AdS wormhole solution violates the null energy condition so that it indeed satisfies the criterion for being wormholes. In the Appendix, detailed computations of KK reduction and Ellis-Bronnikov wormhole are presented.

\section{Correspondence via Kaluza-Klein Reduction}
In this section, we extend the AdS/Ricci-flat correspondence to include an additional axion field.\cite{KKskenderis2, LuMa} The inclusion of the axion field allows us to later explore the connections between wormholes with cosmological constant and wormholes in flat space, thus gaining access to exact results of the otherwise impregnable problems.   
Let's start by introducing the correspondence. One of the theories is the $\hat{D}$-dimensional Einstein gravity with an axion field (phantom field) $\chi$,
\begin{equation}
    \hat{S} = \frac{1}{16 \pi \hat{G}} \int \dd^{\hat{D}} x \hat{\mathcal{L}}_{\hat{D}}
\end{equation} where the Lagrangian density $\hat{\mathcal{L}}$ is simply the Einstein-Hilbert one
\begin{equation} 
    \hat{\mathcal{L}}_{\hat{D}} = \sqrt{-\hat{g}} \Big( \hat{R} + \big( \partial \chi \big)^2 \Big)
\end{equation} 
The correspondence is obtained by performing spherical reduction for this  theory and then torus reduction for the other one. For spherical reduction, the ansatz is given by,
\begin{equation}
  \label{ansatz}
    \dd \hat{s}^2_{\hat{D}} = e^{2\alpha\phi_1} \dd s_d^2 + e^{2\beta \phi_2} \dd \Omega_n^2
\end{equation} where $\hat{D} = d+n$. This reduction is what we call "diagonal" since it does not mix any off-diagonal elements in the metric tensor. One can regard the reduction as removing the fiber part of the bundle, keeping only the theory on the base manifold. This we can do as the dimension of the fibers are extremely small as compared to the base. 
\begin{equation}
    \mathcal{L}_d = \ell^n \sqrt{-g}  X^{\hat{D}-2} \Big(  R +   \big( \hat{D} - 1 \big) \big( \hat{D} - 2 \big) \big( \partial X \big)^2 X^{-2}   + \frac{n(n-1)}{\ell^2}  + \frac{1}{2} \big( \partial \chi \big)^2 \Big)
\end{equation} where the term $n(n-1)/\ell^2$ is the curvature of n-sphere. This theory has the Newton's constant $G_{d} = \ell^{-n} \hat{G} $. \\
The other theory we consider in the correspondence is 
\begin{eqnarray}
    \tilde{S} &=& \frac{1}{16\pi \tilde{G}} \int \dd^{\tilde{D}} \tilde{\mathcal{L}}_{\tilde{D}} \nonumber \\ 
    \tilde{\mathcal{L}}_{\tilde{D}} &=& \sqrt{-\tilde{g}} \Big( \tilde{R} - 2\Lambda + \frac{1}{2} \big( \partial \tilde{\chi} \big)^2 \Big)
\end{eqnarray} For the torus reduction, we use the metric ansatz,
\begin{equation}
    \dd \tilde{s}^2_{\tilde{D}} = \dd \tilde{s}^2_d  +  Y^2 \dd s^2_{\mathcal{T}^Q}
\end{equation}where $\tilde{D} = d + Q$. In the reduction, the field $\chi$ is directly reduces from higher dimension to the low dimension as a scalar field\cite{KKSugra},
\begin{equation}
    \mathcal{L}'_d = \mathcal{V}_Q \sqrt{-g} Y^Q \Big(  R + Q(Q - 1) \big( \partial Y \big)^2 Y^{-2}  + \frac{\Tilde{n}(\Tilde{n} - 1)}{\Tilde{\ell}^2}  +  \frac{1}{2} \big( \partial \chi \big)^2  \Big)
\end{equation} where $\mathcal{V}_Q $ is the volume of the torus, thus giving us the Newton's constant $\Tilde{G}_{d} = \mathcal{V}_Q^{-1} \Tilde{G} $.  Thus we found that the two theories match at lower dimension, provided that we identify the parameter in the following way,
\begin{equation}
    X = Y^{-1}~~,~~~ \ell = \Tilde{\ell}~~,~~~ \ell^n \tilde{G} = \mathcal{V}_Q \hat{G}
\end{equation} with the dimensions of the two theories related by 
\begin{equation}
    n \leftrightarrow \Tilde{n} ~~,~~~ Q \leftrightarrow - \big( \hat{D} -2 \big)
\end{equation}We note that the mapping of the dimension parameters are represented by the $\leftrightarrow$ rather than equality. Because we are mapping one theory with positive dimensions to another one with negative dimensions. Implementing this correspondence, we are able to find exact solutions for one theory by the dictionary as we demonstrate in the next section, providing that we know the solution for the other one.

\section{Mapping of the Solutions of Ellis Wormhole to AdS Ellis Wormhole in General Dimensions}
In this section, we demonstrate in that some special wormholes in higher dimensions with cosmological constant could be solved exactly using the mapping we obtained in the previous section. We first present the exact solution of the well-known Ellis-Bronnikov wormholes in higher dimensions, i.e. where we set $f(r) = 1$ in the metric of the wormholes.\cite{torii} Then we apply the dictionary to extract the exact solution of a special class of wormholes with cosmological constant by comparing the overall factor in the reduced Lagrangian from the two wormhole theories. We shall start with the wormhole in Ricci-flat theory and find the AdS wormhole solution by the dictionary. Recall that we have for Ellis wormhole, the action,
\begin{equation}
  S = \frac{1}{16\pi G} \int \dd^n x \sqrt{-g} \Big( R - \big( \nabla \chi \big)^2 \Big)
\end{equation} We are working with the metric ansatz as for the hat-theory in the previous section which we dimensionally reduced it on sphere. For the $\hat{D} = p+2 + n$ dimensional theory, the metric becomes,
\begin{equation}
    \dd \hat{s}^2_{\hat{D}} = -f(r) \dd t^2 + \frac{\dd r^2}{f(r)} + \rho^2(r) \dd \Omega^2_n
    = \frac{\rho^2}{\ell^2} \Big( \dd s^2_d + \ell^2 \dd \Omega^2_n  \Big)
\end{equation} Note that $p=0$ and $\hat{D} = 2+n$. The functions $f(r)$ and $\rho(r)$ satisfy the equations of motion \cite{torii},
\begin{eqnarray}
     -\frac{1}{2} \chi'^2 &=& (\hat{D} - 2) \frac{\rho'}{\rho} \Big( \frac{f'}{f} + 
    (\hat{D} - 3) \frac{\rho'}{\rho} \Big)  -  (\hat{D} - 2) ( \hat{D} - 3) \frac{1}{ f \rho^2}  \\
    \frac{1}{2} \chi'^2 &=& ( \hat{D} - 2) \frac{\rho''}{\rho}  \\
    \chi' &=& \frac{C}{f \rho^{\hat{D} - 2}}
\end{eqnarray} We find the integration constant 
$ C^2 = (\hat{D}-2)(\hat{D}-3)a^{2(\hat{D}-3)}$. Thus we find the solution to the equations of motion as
\begin{eqnarray}
     f(r) &\equiv& 1   \\
     \rho'(r) &=& \sqrt{1 - \Big( \frac{a}{\rho} \Big)^{2(\hat{D}-3)}}  \\
     \chi(r) &=& \sqrt{2(\hat{D} -2)(\hat{D} -3)} a^{\hat{D}-3} \int \frac{\dd r}{\rho(r)^{\hat{D}-2}}
\end{eqnarray} We've set $f(r) = 1$ for this particular class of wormhole solution to better illustrate the structure of the dual theory. 
Now, in order to see the wormhole structure of this theory, we note that 
\begin{eqnarray}
    \frac{\dd \rho}{\dd r} &=& \sqrt{1-\big( a/ \rho \big)^{2(\hat{D} - 3)} }  \nonumber \\
    \dd r &=& \frac{\dd \rho}{\sqrt{1-\big( a/ \rho \big)^{2(\hat{D} - 3)} }}  \nonumber \\
    \dd r^2 &=& \frac{\dd \rho^2 }{ 1-\big( a/ \rho \big)^{2(\hat{D} - 3)} } 
\end{eqnarray} This we can do as it is a mere coordinate transformation.
After swapping the variables $\rho$ and $r$, we can rewrite the $\dd \hat{S}^2$ metric as 
\begin{equation}
  \label{Eliis}
    \dd \hat{s}^2 = - \dd t^2 +  \frac{\dd \rho^2 }{ 1-\big( a/ \rho \big)^{2(\hat{D} - 3)} } + \rho^2 \dd \Omega^2_n 
    \end{equation} 

The metric of the dimensionally reduced theory is given by
\begin{equation}
    \dd s^2_2 = \frac{1}{X^2} \Big( -f \dd t^2 + \frac{\dd r^2}{f} \Big)
    = \frac{\ell^2}{\rho^2} \Big( -f \dd t^2 + \frac{\dd r^2}{f} \Big)
\end{equation} So apply the mapping, we find that 
\begin{equation}
    \dd \Tilde{s}^2_2 = Y^2 \Big( -\Tilde{f} \dd t^2 + \frac{\dd r^2}{\Tilde{f}} \Big)
                      = \frac{\Tilde{\ell}^2}{\rho^2} \Big( - \Tilde{f} \dd t^2 + \frac{\dd r^2}{\Tilde{f}} \Big)
\end{equation} From here, we recover the full Tilde-theory by lifting the reduced metric back to higher dimensions. Then the $\Tilde{D} = Q + 2$ dimensional metric becomes
\begin{equation}
    \dd \Tilde{s}^2_{\Tilde{D}} = \frac{\Tilde{\ell}^2}{\rho^2} \Big( -\Tilde{f} \dd t^2 + \frac{\dd r^2}{\Tilde{f}}  + \dd y^j \dd y^j \Big) 
\end{equation} where the functions $\Tilde{f}(r)$ and $\rho(r)$ should satisfy the equations,
\begin{eqnarray}
     \frac{1}{2} \chi'^2 &=&  Q~\frac{\rho'}{\rho} \Big( \frac{\Tilde{f}'}{\Tilde{f}}  - \big( Q + 1 \big) \frac{\rho'}{\rho} \Big)  + Q \big(Q + 1\big)   \frac{1}{ \Tilde{f} \rho^2}  \\
    - \frac{1}{2} \chi'^2 &=& Q~ \frac{\rho''}{\rho}  \\
    \chi' &=& \frac{C\rho^q}{ \Tilde{f} } 
\end{eqnarray} where we used the identification we discussed in the previous section, 
\begin{equation}
    Q \leftrightarrow - \Big( \hat{D} -2 \Big)
\end{equation}
Again we can express $ r $ in terms of $\rho$ instead, 
\begin{equation}
     \frac{\dd \rho}{\dd r} = \sqrt{1-\big( a/ \rho \big)^{-2(Q + 1)}}  \Rightarrow  \dd r = \frac{\dd \rho}{\sqrt{1-\big( \rho /a \big)^{2(Q + 1)}} } 
\end{equation} Putting this back to the $\tilde{D} = Q + 2$ dimensional metric, We can rewrite it as 
\begin{equation}
  \label{AdS}
    \dd \Tilde{s}_{\Tilde{D}}^2 = \frac{\ell^2}{\rho^2}\Big(  -\dd t^2 + \frac{\dd \rho^2}{1-\big( \rho /a \big)^{2(q+1)}} + \dd y^j \dd y^j \Big)
\end{equation} and that
\begin{eqnarray}
    \chi(\rho) &=& \sqrt{2Q(Q+1)} a^{-( Q + 1)}  \int  \frac{\rho^Q \dd \rho}{\sqrt{1-\big( \frac{\rho}{a} \big)^{2(Q+1)} }}  \nonumber \\
    &=& \sqrt{\frac{2Q}{Q+1}}  \arcsin{\frac{\rho^{Q+1}}{a^{Q+1}}}
\end{eqnarray}
We show that in general the solution obtained via the correspondence is wormhole solution in section 6.

\section{Ellis-Bronnikov Wormhole in general Dimensions}
Now we proceed to the exact solution of AdS wormhole in general dimensions. For n-dimensional wormhole in Ricci flat space-time, it was found that with specific metric ansatz \cite{nozawa I}, one can obtain the generic solution:
\begin{equation}
  \label{metric}
    \dd s^2 = -F(r)^{-2} \dd t^2 + F(r)^{2/(n-3)} G(r)^{-(n-4)/(n-3)} \Big( \dd r^2 + G(r)\gamma_{ij}(z) \dd z^i \dd z^j  \Big)
\end{equation} where the functions $F(r)$ and $G(r)$ satisfy:
\begin{eqnarray}
  \label{eom1}
    \frac{F(r)''}{F(r)} - \frac{F(r)'}{F(r)} + \frac{F(r)'}{F(r)}\frac{G(r)'}{G(r)} &=& 0   \\
    \label{eom2}
    \frac{F(r)'^2}{F^2(r)} - \frac{1}{4} \frac{G(r)'^2}{G^2(r)} + \frac{(n-3)^2}{G(r)} - \frac{(n - 3)}{2(n - 2)} \frac{ C^2 }{G^2(r)}  &=& 0 
\end{eqnarray} we also have a master equation for $G(r)$ from $E^r_{~r} + E^i_{~i} = 0$,
\begin{equation}
    G(r)'' - 2\big( n - 3 \big)^2 = 0
\end{equation} 
The general solution of these equations, in the case of the Ellis-Bronnikov class solution of this metric, are given by,
\begin{eqnarray}
    \dd s^2 &=& -\frac{1}{F(\bar{r})^2} \dd \bar{t}^2 + F(\bar{r})^{\frac{2}{n-3}}G(\bar{r})^{-\frac{n-4}{n-3}}  \bigg(  \dd \bar{r}^2 + G(\bar{r}) \dd \Omega^2 \bigg)   \nonumber \\
    F(\bar{r}) &=& F_0 \exp \bigg( \beta \arctan{\Big( \frac{\bar{r}}{R} \Big) } \bigg)  \\
    G(\bar{r}) &=& \big( n-3 \big)^2 \Big( \bar{r}^2 + R^2 \Big)   \nonumber 
\end{eqnarray} where we have relabeled  the original coordinates $t$ and $r$ as $\bar{t}$ and $\bar{r}$. Putting all the pieces together, we can write the metric explicitly as
\begin{eqnarray}
    \dd s^2 &=& - e^{-2 \beta \arctan{ ( \bar{r}/R ) } } \frac{\dd \bar{t}^2}{F_0^2}    \\
            &+&  e^{ 2 \beta \arctan{ ( \bar{r}/R )} / (n-3) }  F_0^{2/(n-3)} \big( ( n-3 )^2 ( \bar{r}^2 + R^2 ) \big)^{-(n-4)/(n-3)}  \bigg(  \dd \bar{r}^2 +  ( n-3 )^2 \big( \bar{r}^2 + R^2  \big) \dd \Omega^2 \bigg)   \nonumber 
\end{eqnarray} We arrive at the following metric after redefining the variables $\bar{t}$ and $\bar{r}$ and defining $U(r)$ and $V(r)$ as in Appendix B,
\begin{equation}
     \dd s^2 = -e^{-2\beta U(r)} \dd t^2 + e^{2\beta U(r)/(n-3)} V(r)^{1/(n-3)} \bigg( \frac{\dd r^2 }{V(r)} + r^2 \dd \Omega^2 \bigg) 
\end{equation} 

This solution reduces to the desired wormhole solution that we derived in previous section in the $f \equiv 1$ case. To see this, we first take the following coordinate transformation, 
\begin{equation}
    r = x \qty\Big( 1 - \frac{M^2}{16 x^{2(n-3)}} )^{1/(n-3)}
\end{equation} so the metric can be further rewrite in the form,
\begin{equation}
  \label{metricEB}
    \dd s^2 = - e^{-2\beta \hat{U}(x)} \dd t^2 + e^{2\beta \hat{U}(x)/(n-3)}\qty\Big( 1 + \frac{M^2}{16 x^{2(n-3)}} )^{2/(n-3)} \qty\Big( \dd x^2 + x^2 \dd \Omega^2 )
\end{equation}  We can check explicitly, in 4 dimension, that this metric indeed coincides with the Ellis wormhole with $f \equiv 1 $, or equivalently, $ \beta = 0$. In order to see this, suppose we set $n = 4$, and let $a \equiv \frac{M^2}{4}$, we define another new coordinate $\rho$ as the following,
\begin{equation}
     \rho^2 \equiv x^2 \Big( 1 + \frac{M^2}{16 x^{2(\hat{D} - 3)}} \Big)^{2(\hat{D} - 3)}
\end{equation}
The metric \eqref{metricEB} thus becomes
\begin{equation}
    \dd \hat{s}^2 = -\dd t^2 + \frac{\dd \rho^2}{1- \frac{M^2}{4} / \rho^{2(\hat{D} - 3)} } + \rho^2 \dd \Omega^2    \nonumber 
\end{equation} We've shown that this solution is equivalent to equation\eqref{Eliis}.

\section{AdS Wormhole in General Dimension}
Finally, we proceed to use the mapping to compute the solutions of tilde theory. We already have the hat theory given as 
\begin{eqnarray}
    \dd \hat{s}^2 &=& -\frac{\dd t^2}{F(r)^2 }  + F(r)^{2/(\hat{D} - 3)}G(r)^{-(\hat{D} -4)/(\hat{D} -3)} \qty\Big( \dd r^2 + G(r) \dd \Omega^2 )   \nonumber \\
    &=& \frac{\rho(r)^2}{\ell^2} \qty\Big( \dd s^2_2 +\ell^2 \dd \Omega^2 ) ,~~~~~~ \rho(r) = F(r)^{1/(\hat{D} -3)} G(r)^{1/2(\hat{D}-3)} \\
    \dd s^2_2 &=& \frac{\ell^2}{\rho(r)^2} \qty( -\frac{1}{F(r)^2} \dd t^2  + F(r)^{2/(\hat{D} -3 )} G(r)^{- (\hat{D}-4)/(\hat{D}-3)}  \dd r^2)    \nonumber  \\
    X^2 &=& \frac{\rho(r)^2}{\ell^2} = \frac{ 1}{\ell^2} F(r)^{2/(\hat{D} -3)} G(r)^{1/(\hat{D}-3)}  \nonumber \\
    F(r) = F_0 e^{\beta U(r)} &,&~~ G(r) = r^{2(\hat{D} - 3)} V(r) ~~,~~ U(r) = \arctan{\qty(\frac{2r^{\hat{D}-3}}{M})} ~~,~~ V(r) = 1 + \frac{M^2}{4r^{2(\hat{D} -3)}}  \nonumber 
\end{eqnarray}
Apply the mapping:
\begin{eqnarray}
    Y^2 &=& \frac{1}{X^2} = \ell^2  \Tilde{F}(r)^{2/(Q + 1)} \Tilde{G}(r)^{1/(Q+1)}   \\ 
    \Tilde{F}(r) = F_0 e^{\beta \Tilde{U}(r)} ~~, ~~  \Tilde{G}(r) &=& \frac{\Tilde{V}(r)}{r^{2(Q+1)} }  ~~, ~~\Tilde{U}(r) = \arctan{\qty( \frac{2}{\Tilde{M} r^{Q+1} } )} ~~,~~ \Tilde{V}(r) = 1 + \frac{\Tilde{M}^2 r^{2(Q+1)}}{4}    \nonumber 
\end{eqnarray} Thus the $\Tilde{D} = 2 + Q$ dimensional metric becomes
\begin{equation}
    \dd \Tilde{s}^2 = Y^2 \qty\Big( -\frac{\dd t^2}{\Tilde{F}(r)^2}  + \Tilde{F}(r)^{-2/(Q+1)} \Tilde{G}(r)^{-(Q+2)/(Q+1)} \dd r^2 + \dd y^j \dd y^j )
\end{equation} 
Lastly, we write the Tilde theory explicitly,
\begin{equation}
    \dd \Tilde{s}^2 = \ell^2  \qty\Big( - e^{-2\beta\Tilde{U}(r)/(Q+1)}  \frac{\Tilde{V}(r)^{1/(Q+1)}}{r^2} \dd t^2     +  \frac{r^{2(Q+1)}}{\Tilde{V}(r)} \dd r^2 + e^{2\beta \Tilde{U}/(Q+1)} \frac{\Tilde{V}^{1/(Q+1)}}{r^2} \dd y^j \dd y^j )
\end{equation} This is the desired wormhole solution in general dimensions. We notice that the dimension parameter $Q$ ranges form negative infinity to positive infinity. It is worth looking at the sign of $Q$ to exam the properties of this solution.

From here, we do similar coordinate transformation as for the hat-theory in the previous section. We take the 
\begin{equation}
    r^2 = x^2\Big( 1 - \frac{M^2}{16} x^{2(Q+1)}  \Big)^{-\frac{2}{Q+1}}
\end{equation} Here let's write out the metric after the above coordinate transformation,
\begin{equation}
    \dd \tilde{s}^2 = \frac{\ell^2 }{ x^2 }  \big( 1 + \frac{M^2}{16} x^{2(Q+1)} \big)^{\frac{2}{Q+1}} \Big( - e^{ -2\beta \tilde{U}/(Q+1)} \dd t^2 + \frac{\dd x^2}{\big( 1+\frac{M^2}{16}x^{2(Q+1)} \big)^{\frac{2}{Q+1}} } + e^{2\beta \tilde{U}/(Q+1)} \dd y^i \dd y^i \Big)
\end{equation}  A further coordinate transformation
\begin{equation}
    \rho^2 = x^2\Big( 1 + \frac{M^2}{16} x^{2(Q+1)} \Big)^{-\frac{2}{Q+1}}
\end{equation} gives us the desired form of solution,
\begin{equation}
    \dd \tilde{s}^2_{\tilde{D}} = \frac{\ell^2}{\rho^2} \bigg( - e^{ -2\beta \tilde{U}/(Q+1)} \dd t^2 + \frac{\dd \rho^2}{1 - \rho^{2(Q+1)}/a}  + e^{2\beta \tilde{U}/(Q+1)} \dd y^i \dd y^i \bigg)
\end{equation}

For the case $\beta = 0 $, the solution takes the form 
\begin{equation}
    \dd \Tilde{s}^2_{\Tilde{D}} = \frac{\ell^2}{\rho^2} \bigg( - \dd t^2 + \frac{\dd \rho^2}{1 - \rho^{2(Q+1)}/a} + \dd y^i \dd y^i \bigg)
\end{equation} This is the same as the solution \eqref{AdS} we found in section 3.

\section{Null Energy Condition}
It is known that in order for a solution to be a wormhole, it needs to violate the null energy condition \cite{NullEnergy} 
\begin{equation}
     T_{\mu\nu} n^{\mu} n^{\nu} \ge 0 
\end{equation}We use the null vector 
\begin{equation}
    n_{\mu} = \big( g_{tt}, \sqrt{g_{tt} g_{\rho\rho}} , 0, 0 \big)
\end{equation} whose covariant conterpart is given by
\begin{equation}
    n^{\mu}= \big( 1, \sqrt{g_{tt}/g_{\rho \rho} }  , 0 , 0  \big)
\end{equation} The null energy condition takes the form
\begin{equation}
    T_{\mu\nu} n^{\mu} n^{\nu} = T^t_t n^t n_t + T^{\rho}_{\rho} n^{\rho} n_{\rho} = -g_{tt} \Big( -T^t_t + T^{\rho}_{\rho} \Big)
\end{equation} Here we insert some steps of computing this value by looking at the Einstein equations' components and metric elements in the Mathematica files we made.
\begin{eqnarray}
    T_{\mu\nu} n^{\mu} n^{\nu} &=& T_{tt}- g_{tt}T^{\rho}_{\rho} \nonumber \\
               &=& - \frac{a+ Q\rho^{2+2Q} + a\ell^2 \Lambda}{a\rho^2} + \frac{\rho^2}{\ell^2} \Big( \frac{1}{\rho^2} + \frac{a\ell^2 \Lambda}{a\rho^2 - \rho^{4+2Q}} \Big) \nonumber \\
    &=& \frac{1}{\ell^2} - \frac{1}{\rho^2} - \frac{\ell^2 \Lambda}{\rho^2} - \frac{Q \rho^{2Q}}{a} + \frac{a\Lambda}{a - \rho^{2+2Q}} < 0
\end{eqnarray} This theory violates the null energy condition and thus indeed is a wormhole solution.




\section{Conclusion and Summary}
The wormholes with cosmological constant in higher dimensions are in general hard to solve directly. We found the general exact solution of wormholes with cosmological constant in higher dimensions. This was achieved by implementing the mapping method that we learnt from Kaluza-Klein reduction of two different theories which were identified after dimensionally reducing to the same lower theory. We also studied the property of such solution and demonstrated that they are indeed good wormholes. It is of further interest if one could try to find a correspondence more general that maps different theories with gauge theories involved, thus solving more profound problems that were otherwise impossible to tackle with.

\section{Acknowledgement}
We thank Professor Hong Lu for his advice and helpful discussion on the inclusion of axion fields in AdS-Ricci flat correspondence. We are deeply grateful for his support and his warm hospitality at Tianjin University where author was a visitor during summer 2021.

\newpage

\appendix

\section{Kaluza-Klein Reduction}
In this Appendix, we focus on deriving the reduced Lagrangian in lower dimension. In section II, we have the metric ansatz \eqref{ansatz} for spherical reduction,
\begin{equation}
    \dd \hat{s}^2_{\hat{D}} = e^{2\alpha\phi_1} \dd s_d^2 + e^{2\beta \phi_2} \dd \Omega_n^2
\end{equation}  
We choose the following vielbien for this reduction,
\begin{equation}
    \hat{\text{E}}^a  = e^{\alpha\phi_1} \text{E}^a  ~~~,~~~~~
    \hat{\text{E}}^i  = e^{\beta\phi_2} \text{E}^i
\end{equation}Upon taking exterior derivative, 
\begin{eqnarray}
    \dd \hat{\text{E}}^a &=& -\hat{\omega}^a_{~b} \wedge \hat{\text{E}}^b - \hat{\omega}^a_{~j} \wedge \hat{\text{E}}^j\nonumber \\
       &=& \dd \Big( e^{\alpha\phi_1} \text{E}^a \Big) = \dd \big( e^{\alpha \phi_1} \big) \wedge \text{E}^a + e^{\alpha\phi_1} \dd \text{E}^a  \nonumber \\
       &=& \alpha e^{\alpha\phi_1} \partial_b \phi_1 \text{E}^b \wedge \text{E}^a + e^{\alpha \phi_1} \big( -\omega^a_{~b} \wedge \text{E}^b\big) \\
       &=& \alpha e^{-\alpha \phi_1} \partial_b \phi_1 \hat{\text{E}}^b \wedge \hat{\text{E}}^a - \omega^a_{~b} \wedge \hat{\text{E}}^b   \nonumber 
\end{eqnarray} where we used Cartan's first structural equation in the first line and the fact that $\omega^a_j = 0$ for our diagonal reduction ansatz in the third line. So we can read off the spin connection 1-form as
\begin{equation}
    \hat{\omega}^a_{~b} = \omega^a_{~b} +  \alpha e^{-\alpha \phi_1} \Big( \partial_b \phi_1 \hat{\text{E}}^a - \partial^a \phi_1 \hat{\text{E}}^c \delta_{cb} \Big) 
\end{equation} We obtain all the spin connections in similar way and list them here,
\begin{eqnarray}
\hat{\omega}^a_{~b} &=& \omega^a_{~b} +  \alpha e^{-\alpha \phi_1} \Big( \partial_b \phi_1 \hat{\text{E}}^a - \partial^a \phi_1 \hat{\text{E}}^c \delta_{cb} \Big)   \nonumber \\
    \hat{\omega}^i_{~j} &=& \omega^i_{~j}    \nonumber \\
    \hat{\omega}^i_{~c} &=& \beta e^{- \alpha \phi_1} \partial_c \phi_2 \hat{\text{E}}^i
\end{eqnarray} The curvature 2-forms $\hat{\Omega}$'s can be computed by using the spin connection 1-forms we found. To proceed, we recall the definition
\begin{equation}
    \hat{\Omega}^a_{~b} = \dd \hat{\omega}^a_{~b} + \hat{\omega}^a_{~c} \wedge \hat{\omega}^c_{~b} + \hat{\omega}^a_{~j} \wedge \hat{\omega}^j_{~b}
\end{equation}
The first term is given by,
\begin{eqnarray}
  \label{7}
    \dd \hat{\omega}^a_{~b} &=& \dd \omega^a_{~b} + \alpha \big( -\alpha e^{-\alpha\phi_1} \big) \big( \dd \phi_1 \big) \wedge \big( \partial_b \phi_1 \hat{\text{E}}^a - \partial^a \phi_1 \hat{\text{E}}^c \delta_{cb}  \big) \nonumber \\
    && + \alpha e^{-\alpha \phi_1} \dd \Big( \partial_b \phi_1 \hat{\text{E}}^a - \partial^a \phi_1 \hat{\text{E}}^c \delta_{cb} \Big) \nonumber \\
    &=& \dd \omega^a_{~b} - \alpha^2 e^{-2\alpha\phi_1} \partial_c \phi_1 \hat{\text{E}}^c \wedge \Big( \partial_b \phi_1 \hat{\text{E}}^a - \partial^a \phi_1 \hat{\text{E}}^d \delta_{db}   \Big)  \\
    && + \alpha e^{-2\alpha \phi_1} \Big(   \partial_c \partial_b \phi_1 \hat{\text{E}}^c \wedge \hat{\text{E}}^a - \partial_c \partial^a \phi_1 \hat{\text{E}}^c \wedge \hat{\text{E}}^d \delta_{db} \Big)  + \alpha e^{- \alpha \phi_1} \Big( \partial_b \phi_1 \dd \hat{\text{E}}^a - \partial^a \phi_1 \dd \hat{\text{E}}^d \delta_{db}  \Big)   \nonumber 
\end{eqnarray} Note that the terms in the last bracket involves exterior derivative of $\hat{\text{E}}$'s, we want to show in detail the computation of these terms.
\begin{eqnarray}
  \label{8}
    \alpha e^{- \alpha \phi_1} \partial_b \phi_1 \dd \hat{\text{E}}^a &=& \alpha e^{- \alpha \phi_1} \partial_b \phi_1 \Big( -\alpha e^{-\alpha\phi_1} \partial_c \phi_1 \hat{\text{E}}^a \wedge \hat{\text{E}}^c  - \omega^a_{~c} \wedge \hat{\text{E}}^c \Big)  \nonumber \\
    &=& \alpha^2 e^{-2\alpha \phi_1} \partial_b \phi_1 \partial_c \phi_1 \hat{\text{E}}^c \wedge \hat{\text{E}}^a  -  \alpha e^{-\alpha\phi_1}  \partial_b \phi_1 \omega^a_{~c} \wedge \hat{\text{E}}^c
\end{eqnarray} 

Plug equation\eqref{8} back into equation\eqref{7}, we find that
\begin{eqnarray}
    \dd \hat{\omega}^a_{~b} &=& \dd \omega^a_{~b} - \alpha^2 e^{-2\alpha\phi_1} \partial_c \phi_1   \partial_b \phi_1 \hat{\text{E}}^c \wedge \hat{\text{E}}^a + \alpha^2 e^{-2\alpha\phi_1} \partial_c \phi_1 \partial^a \phi_1 \hat{\text{E}}^c \wedge \hat{\text{E}}^d \delta_{db}  \nonumber  \\
    && + \alpha e^{-2 \alpha \phi_1} \Big(    \partial_c \partial_b \phi_1 \hat{\text{E}}^c \wedge \hat{\text{E}}^a - \partial_c \partial^a \phi_1 \hat{\text{E}}^c \wedge \hat{\text{E}}^d \delta_{db}   \Big)   \nonumber  \\
    && + \alpha^2 e^{-2\alpha \phi_1} \partial_b \phi_1 \partial_c \phi_1 \hat{\text{E}}^c \wedge \hat{\text{E}}^a  -  \alpha e^{-\alpha\phi_1} \partial_b \phi_1 \omega^a_{~c} \wedge \hat{\text{E}}^c   \nonumber \\
    && - \alpha^2 e^{-2\alpha \phi_1} \partial^a \phi_1 \partial_c \phi_1 \hat{\text{E}}^c \wedge \hat{\text{E}}^d \delta_{db}   +   \alpha e^{-\alpha\phi_1}  \partial^a \phi_1 \omega^d_{~f} \wedge \hat{\text{E}}^f \delta_{db} \nonumber   \\
    &=& \dd \omega^a_{~b}  - \alpha e^{-2 \alpha \phi_1} \Big(   \partial_c \partial_b \phi_1  \hat{\text{E}}^a  +  \omega^a_{~cf} \partial_b \phi_1 \hat{\text{E}}^f \Big) \wedge \hat{\text{E}}^c   \nonumber \\
    && + \alpha e^{-2 \alpha \phi_1} \Big(  \partial_c \partial^a \phi_1  \hat{\text{E}}^d   +   \omega^d_{~cf} \partial^a \phi_1  \hat{\text{E}}^f \Big) \delta_{db} \wedge \hat{\text{E}}^c   \nonumber \\    
    &=&   \dd \omega^a_{~b} + \alpha e^{-2 \alpha \phi_1} \Big(  \nabla_c \nabla^a \phi_1 \hat{\text{E}}^d \delta_{db}  
    - \nabla_b \nabla_c \phi_1 \hat{\text{E}}^a   \Big)  \wedge \hat{\text{E}}^c
\end{eqnarray} where in the last line, we used the definition,
\begin{eqnarray}
    \nabla_a V^b &\equiv& \partial_a V^b + \omega_{a~~c}^{~b} V^c     \nonumber \\
    \nabla_a V_b &\equiv& \partial_a V_b - \omega_{a~~b}^{~c} V_c
\end{eqnarray} the $\omega_{a~~b}^{~c}$ comes from the definition 
\begin{equation}
    \omega^a_{~b} \equiv \omega^a_{~bc} \text{E}^c
\end{equation}Then we can compute the curvature 2-form $\hat{\Omega}^a_{~b}$,
\begin{eqnarray}
    \hat{\Omega}^a_{~b} &=& \Big( \dd \omega^a_{~b}  +  \omega^a_{~c} \wedge  \omega^c_{~b} \Big) + \alpha e^{-2 \alpha \phi_1} \Big(  \nabla_c \nabla^a \phi_1 \hat{\text{E}}^d \delta_{db}  
    - \nabla_b \nabla_c \phi_1 \hat{\text{E}}^a   \Big)  \wedge \hat{\text{E}}^c  \nonumber \\
    &&   +   \alpha^2 e^{-2 \alpha \phi_1} \Big(  \partial_c \phi_1 \hat{\text{E}}^a  - \partial^a \phi_1 \hat{\text{E}}^d \delta_{dc}  \Big) \wedge \Big(   \partial_b \phi_1 \hat{\text{E}}^c  - \partial^c \phi_1 \hat{\text{E}}^f \delta_{fb}    \Big)  \nonumber \\
    && +   \alpha e^{- \alpha \phi_1} \Big[  \omega^a_{~c} \wedge   \Big(   \partial_b \phi_1 \hat{\text{E}}^c  - \partial^c \phi_1 \hat{\text{E}}^f \delta_{fb}    \Big)    
    +  \Big(  \partial_c \phi_1 \hat{\text{E}}^a  - \partial^a \phi_1 \hat{\text{E}}^d \delta_{dc}  \Big) \wedge  \omega^c_{~b}   \Big]  \nonumber \\
    && - \beta^2 e^{-2 \alpha \phi_1} \partial_a \phi_2 \hat{\text{E}}^k \wedge \partial_b \phi_2 \hat{\text{E}}^j \delta_{kj}     \nonumber \\
    &=& \Omega^a_{~b}   + \alpha e^{-2 \alpha \phi_1} \Big(  \nabla_c \nabla^a \phi_1 \hat{\text{E}}^d \delta_{db}  
    - \nabla_b \nabla_c \phi_1 \hat{\text{E}}^a   \Big)  \wedge \hat{\text{E}}^c  \nonumber \\
    && + \alpha^2 e^{-2 \alpha \phi_1} \Big(  \partial_c \phi_1 \hat{\text{E}}^a  - \partial^a \phi_1 \hat{\text{E}}^d \delta_{dc}  \Big) \wedge \Big(   \partial_b \phi_1 \hat{\text{E}}^c  - \partial^c \phi_1 \hat{\text{E}}^f \delta_{fb}    \Big) 
\end{eqnarray} where we noted that 
$$\hat{\text{E}}^k \wedge \hat{\text{E}}^j \delta_{kj} = \hat{\text{E}}^j \wedge \hat{\text{E}}^j = 0$$ We further find that
\begin{eqnarray}
     \hat{\Omega}^a_{~b} &=& \Omega^a_{~b}   + \alpha e^{-2 \alpha \phi_1} \Big(  \nabla_c \nabla^a \phi_1 \hat{\text{E}}^d \delta_{db}  
    - \nabla_b \nabla_c \phi_1 \hat{\text{E}}^a   \Big)  \wedge \hat{\text{E}}^c  \nonumber \\
    && - \alpha^2 e^{-2 \alpha \phi_1} \big( \partial \phi_1  \big)^2 \hat{\text{E}}^a \wedge \hat{\text{E}}^f \delta_{fb}    \nonumber \\
    && + \alpha^2 e^{-2 \alpha \phi_1} \partial_c \phi_1 \Big( \partial_b \phi_1 \hat{\text{E}}^a 
       - \partial^a \phi_1  \hat{\text{E}}^f \delta_{fb}  \Big)  \wedge \hat{\text{E}}^c
\end{eqnarray} We list here all the curvature 2-forms,
\begin{eqnarray}
  \label{9}
    \hat{\Omega}^a_{~b} &=& \Omega^a_{~b}   + \alpha e^{-2 \alpha \phi_1} \Big(  \nabla_c \nabla^a \phi_1 \hat{\text{E}}^d \delta_{db}  
    - \nabla_b \nabla_c \phi_1 \hat{\text{E}}^a   \Big)  \wedge \hat{\text{E}}^c  \nonumber \\
    && - \alpha^2 e^{-2 \alpha \phi_1} \big( \partial \phi_1  \big)^2 \hat{\text{E}}^a \wedge \hat{\text{E}}^f \delta_{fb}    \nonumber \\
    && + \alpha^2 e^{-2 \alpha \phi_1} \partial_c \phi_1 \Big( \partial_b \phi_1 \hat{\text{E}}^a 
       - \partial^a \phi_1  \hat{\text{E}}^f \delta_{fb}  \Big)  \wedge \hat{\text{E}}^c  \nonumber \\
     \hat{\Omega}^i_{~j} &=&  \Omega^i_{~j}  -  \beta^2 e^{-2 \alpha \phi_1} \big( \partial \phi_2 \big)^2 \hat{\text{E}}^i \wedge \hat{\text{E}}^k \delta_{kj}   \\
     \hat{\Omega}^i_{~b} &=& -\alpha \beta e^{-2 \alpha \phi_1} 
     \Big( \partial_c \phi_1 \partial_b \phi_2 \hat{\text{E}}^c   
           + \partial_c \phi_2 \partial_b \phi_1 \hat{\text{E}}^c  
           - \partial_a \phi_2 \partial^a \phi_1 \hat{\text{E}}^c \delta_{cb} \Big) \wedge \hat{\text{E}}^i
     \nonumber \\
     && + \beta e^{-2 \alpha \phi_1} \Big( \beta \partial_b \partial_c \phi_2 + \nabla_b \nabla_c \phi_2  \Big) \hat{\text{E}}^c \wedge \hat{\text{E}}^i    \nonumber 
\end{eqnarray}
Now we use the following equation to read off the Riemann curvature
\begin{equation}
    \hat{\Omega}^A_{~B} = \frac{1}{2} \hat{R}^A_{~BMN}  \hat{\text{E}}^M \wedge \hat{\text{E}}^N
\end{equation} where $A, B, M, N \in \{ 1, \dots, d, d+1, \dots, d+n \}$.
We first look at 
\begin{eqnarray}
     \hat{\Omega}^a_{~b} &=& \frac{1}{2} \hat{R}^a_{~bMN} \hat{\text{E}}^M \wedge \hat{\text{E}}^N  \nonumber \\
     &=&  \frac{1}{2} \hat{R}^a_{~bcd} \hat{\text{E}}^c \wedge \hat{\text{E}}^d  
       +  \frac{1}{2} \hat{R}^a_{~bij} \hat{\text{E}}^i \wedge \hat{\text{E}}^j
       +  \frac{1}{2} \hat{R}^a_{~bcj} \hat{\text{E}}^c \wedge \hat{\text{E}}^j
\end{eqnarray} 
compare this with equations \eqref{9} and note that
\begin{equation}
     \Omega^a_{~b} = \frac{1}{2} R^a_{~bMN} \text{E}^M \wedge \text{E}^N 
                   = \frac{1}{2} R^a_{~bcd} \text{E}^c \wedge \text{E}^d
                   = e^{-2 \alpha \phi_1 }\frac{1}{2} R^a_{~bcd} \hat{\text{E}}^c \wedge \hat{\text{E}}^d
\end{equation} We then read off 
\begin{eqnarray}
     \hat{R}^a_{~bcd} &=& e^{-2 \alpha \phi_1} R^a_{~bcd} -  \alpha^2 e^{-2 \alpha \phi_1} \big( \partial \phi_1 \big)^2 \big(  \delta^a_{~c} \delta_{bd}  -  \delta^a_{~d} \delta_{bc}  \big)      \\
     && +  \alpha e^{-2 \alpha \phi_1} \Big( \nabla_b \nabla_c \phi_1 \delta^a_{~d} - \nabla_b \nabla_d \phi_1 \delta^a_{~c}
     - \nabla^a \nabla_c \phi_1 \delta_{bd} + \nabla^a \nabla_d \phi_1 \delta_{bc} \Big)    \nonumber   \\
     && - \alpha^2 e^{-2 \alpha \phi_1} \Big(  \partial_b \phi_1 \partial_c \phi_1 \delta^a_{~d}  -  \partial_b \phi_1 \partial_d \phi_1 \delta^a_{~c}
     - \partial^a \phi_1 \partial_c \phi_1 \delta_{bd} +\partial^a \phi_1 \partial_d \phi_1 \delta_{bc} \Big)  \nonumber \\
      \hat{R}^i_{~bjd} &=&  \alpha \beta e^{-2 \alpha \phi_1} 
      \Big( 2 \partial_d \phi_1 \partial_b \phi_2 \delta^i_{~j}  
      + 2 \partial_d \phi_2 \partial_b \phi_1 \delta^i_{~j} 
      - 2 \partial_a \phi_1 \partial^a \phi_2  \delta_{bd} \delta^i_{~j}  \Big)      \\
      && - 2 \beta e^{-2 \alpha \phi_1} \Big( \beta \partial_b \phi_1 \partial_d \phi_2  +  \nabla_b \nabla_d  \phi_2  \Big)  \delta^i_{~j}  \nonumber  \\
      \hat{R}^i_{~jkl} &=&  e^{-2\beta \phi_2} R^i_{~jkl}  -  \beta^2 e^{-2 \alpha \phi_1} \big( \partial \phi_2 \big)^2 \big( \delta^i_{~k} \delta_{jl} - \delta^i_{~l}  \delta_{jk}  \big)               
\end{eqnarray} where we anti-symmetrized the indices to insure the anti-symmetry of Riemann curvature tensor.
Finally, we contract Riemann curvatures,
\begin{eqnarray}
     \hat{R}_{bd} &=& \hat{R}^a_{~bad} + \hat{R}^i_{~bid}  \nonumber   \\
     &=& e^{-2\alpha \phi_1} R_{bd}           \nonumber \\
     && - \big( d - 1 \big) \alpha^2 e^{-2 \alpha\phi_1} \big(  \partial \phi_1 \big)^2 \delta_{bd}  
     -  \alpha e^{-2 \alpha\phi_1} \Big( \big( d - 2 \big) \nabla_b\nabla_d \phi_1  +  \Box \phi_1  \delta_{bd}  \Big)   \nonumber \\
     && + \alpha^2 e^{-2 \alpha \phi_1} \Big( \big( d - 2 \big) \partial_b \phi_1 \partial_d \phi_1   +  \big(  \partial \phi_1 \big)^2  \delta_{bd}  \Big)   \nonumber  \\
     &&  + 2 n \alpha \beta e^{-2 \alpha \phi_1} \Big(  \partial_d \phi_1 \partial_b \phi_2 + \partial_d \phi_2 \partial_b \phi_1 
     - \partial_a \phi_1 \partial^a \phi_2 \delta_{bd}   \Big)       \\
     &&  - 2 n \beta e^{-2\alpha \phi_1}  \Big(  \beta  \partial_b \phi_2 \partial_d \phi_2 + \nabla_b \nabla_d \phi_2  \Big)   \nonumber   \\
     \hat{R}_{jl} &=& e^{-2\beta \phi_2} R_{jl}  -  \big( n - 1 \big) \beta^2 e^{-2\alpha\phi_1} \big( \partial \phi_2  \big)^2 \delta_{jl}   
\end{eqnarray} 
Eventually, we find that 
\begin{eqnarray}
     \hat{R} &=& e^{-2\alpha\phi_1} R + e^{-2\beta \phi_2} R_{\Omega} - n \big( n - 1 \big) \beta^2 e^{-2 \alpha\phi_1} \big( \nabla \phi_2 \big)^2  
     - d \big( d - 1 \big) \alpha^2 e^{-2\alpha\phi_1} \big( \nabla \phi_1  \big)^2    \nonumber \\
     && +  2  \big( d - 1 \big) \alpha e^{-2 \alpha\phi_1} \Big( \Box \phi_1 
        -  \alpha \big( \nabla  \phi_1 \big)^2 \Big)   
        - 2 n \beta e^{-2 \alpha\phi_1} \Big(  \beta  \big( \nabla \phi_2 \big)^2  + \Box \phi_2  \Big) \nonumber  \\
     && + \big( 4n - 2nd \big) \alpha \beta e^{-2 \alpha \phi_1}   \nabla_a \phi_1 \nabla^a \phi_2 
\end{eqnarray}
Now note the metric determinant in lower dimension is given by,
\begin{equation}
    \sqrt{-\hat{g}} = \sqrt{-\big( e^{2\alpha \phi_1} \big)^d \big( e^{2\beta\phi_2} \big)^n g} 
    = e^{d \alpha \phi_1 + n \beta \phi_2} \sqrt{-g} 
\end{equation} Collecting the pieces above, we find the reduced Lagrangian,
\begin{eqnarray}
     \mathcal{L}_d &=& \sqrt{-\hat{g}} \hat{R}    \nonumber  \\
     &=&   \sqrt{-g} \bigg[ e^{-2\alpha\phi_1} R + e^{-2\beta \phi_2} R_{\Omega}  +  2  \big( d - 1 \big) \alpha e^{-2 \alpha\phi_1} \Big( \Box \phi_1 
        -  \alpha \big( \nabla  \phi_1 \big)^2 \Big)  \nonumber  \\
     && ~~~~~~~~~ -  2n \beta e^{-2 \alpha\phi_1} \Big(  \beta  \big( \nabla \phi_2 \big)^2  + \Box \phi_2  \Big)  
     + 2n \big( 2 - d \big)  \alpha \beta e^{-2 \alpha \phi_1}   \nabla_a \phi_1 \nabla^a \phi_2   \nonumber \\
     && ~~~~~~~~~
        - d \big( d - 1 \big) \alpha^2 e^{-2\alpha\phi_1} \big( \nabla \phi_1  \big)^2   
        - n \big( n - 1 \big)  \beta^2 e^{-2 \alpha\phi_1} \big( \nabla \phi_2 \big)^2  
      \bigg]  e^{d \alpha \phi_1 + n \beta \phi_2}  \nonumber   \\
     &=& \sqrt{-g} \bigg[   e^{-2\alpha\phi_1} \bigg( R  +  2  \big( d - 1 \big) \alpha \Big( \Box \phi_1 
        -  \alpha \big( \nabla  \phi_1 \big)^2 \Big)  - 2 n \beta  \Big(  \beta  \big( \nabla \phi_2 \big)^2  + \Box \phi_2  \Big)   \nonumber \\
     && ~~~~~~~~~~~~~~~
        - d \big( d - 1 \big) \alpha^2  \big( \nabla \phi_1  \big)^2  
        - n \big( n - 1 \big) \beta^2  \big( \nabla \phi_2 \big)^2   \nonumber  \\
     && ~~~~~~~~~~~~~~~
        + 2n  \big( d - 2 \big)  \alpha \beta    \nabla_a \phi_1 \nabla^a \phi_2       \bigg)  + e^{-2\beta \phi_2} R_{\Omega} ~~ \bigg] e^{d \alpha \phi_1 + n \beta \phi_2} 
\end{eqnarray} If we set.
\begin{equation}
     \phi_1 = \frac{1}{\alpha} \ln{X}   ~~,~~~~
     \phi_2 = \frac{1}{\beta} \ln \big( X \ell \big)
\end{equation}
We finally arrive 
\begin{equation}
     \mathcal{L}_d = \ell^n \sqrt{-g}  X^{\hat{D}-2} \Big(  R +   \big( \hat{D} - 1 \big) \big( \hat{D} - 2 \big) \big( \partial X \big)^2 X^{-2}   + \frac{n(n-1)}{\ell^2} \Big)
\end{equation} where $\hat{D} = d+n$ and the last term is the curvature of n-sphere.

\section{Solving the Higher Dimensional Wormholes}
Here we derive in detail the solution of Ellis-Bronnikove wormholes. We start with the action 
\begin{equation}
    S = \int \dd^n x \sqrt{-g} \Big( R + \frac{1}{2} \big( \partial \phi \big)^2 \Big)
\end{equation} The equations of motion are given by
\begin{eqnarray}
    0 &=& R_{\mu\nu} - \frac{1}{2} R g_{\mu\nu} + \frac{1}{2} \Big( \partial_{\mu} \phi \partial_{\nu} \phi - \frac{1}{2} \big( \partial \phi \big)^2  \Big)  \equiv E_{\mu\nu}  \\
    0 &=& \Box \phi \equiv \frac{1}{\sqrt{-g}} \partial_{\mu} \Big( \sqrt{-g} g^{\mu\nu} \partial_{\nu} \phi \Big)
\end{eqnarray} Assume that $\phi = \phi(r)$, and use the metric ansatz \eqref{metric}, we find from the last equation that
\begin{equation}
  \label{phi}
    \frac{\dd \phi}{\dd r} = \frac{C}{G(r)}
\end{equation}
Now we use the equation
\begin{equation}
    E^r_{~r} + E^i_{~i} = 0
\end{equation} 
Then we conclude that the master equation for $G(r)$,
\begin{equation}
  \label{master}
    G'' - 2\big( n - 3 \big)^2 = 0
\end{equation} Then for $E^t_{~t}$, and $E^r_{~r}$, we use Mathematica with mathematical induction and find the equation of motion in generic dimension to be,  
\begin{eqnarray}
    E^t_{~t} &=&  F(r)^{\frac{-2}{n-3}} G(r)^{\frac{n-4}{n-3}} \Bigg( \frac{n-2}{8(n - 3)} \bigg( 8 \frac{F''}{F} + 8 \frac{F'G'}{FG}  - 4\frac{F'^2}{F^2}  - \frac{G'^2}{G^2}  +  \frac{4(n - 3)^2 }{G}   \bigg)  -  \frac{1}{4} \frac{C^2}{G^2}   \Bigg)   \nonumber  \\
    E^r_{~r} &=& F(r)^{\frac{-2}{n-3}} G(r)^{\frac{n-4}{n-3}}  \Bigg( \frac{n-2}{8(n - 3)} \bigg( - 4\frac{F'^2}{F^2}  + \frac{G'^2}{G^2}  -  \frac{4(n - 3)^2 }{G}   \bigg)  +  \frac{1}{4} \frac{C^2}{G^2}   \Bigg) 
\end{eqnarray}
Since we have $E_{\mu\nu} = $, then $E^t_{~t} = 0$ and $E^r_{~r} = 0$, so we find
\begin{equation}
    \frac{1}{4}\frac{C^2}{G^2} = \frac{n-2}{8(n - 3)} \bigg( 8 \frac{F''}{F} + 8 \frac{F'G'}{FG}  - 4\frac{F'^2}{F^2}  - \frac{G'^2}{G^2}  +  \frac{4(n - 3)^2 }{G}   \bigg)
\end{equation} Substitute this back into $E^r_{~r}$, we find that
\begin{eqnarray}
  \label{eomA}
    E^r_{~r} &=& F(r)^{\frac{-2}{n-3}} G(r)^{\frac{n-4}{n-3}}  \Bigg( \frac{n-2}{8(n - 3)} \bigg( - 4\frac{F'^2}{F^2}  + \frac{G'^2}{G^2}  -  \frac{4(n - 3)^2 }{G}   \nonumber \\
    && ~~~~~~~~~~~~~~~~~~~~~~~ + 8 \frac{F''}{F} + 8 \frac{F'G'}{FG}  - 4\frac{F'^2}{F^2}  - \frac{G'^2}{G^2}  +  \frac{4(n - 3)^2 }{G}   \bigg)  \Bigg)   \nonumber \\
    0 &=& F(r)^{\frac{-2}{n-3}} G(r)^{\frac{n-4}{n-3}}  \Bigg( \frac{n-2}{8(n - 3)} \bigg(8 \frac{F''}{F}  + 8 \frac{F'G'}{FG}  -  8 \frac{F'^2}{F^2}  \Bigg)  \nonumber  \\
    0 &=& \bigg( \frac{F''}{F}  +  \frac{F'G'}{FG}  -   \frac{F'^2}{F^2}  \Bigg)
\end{eqnarray}
Also from $E^r+{~r} = 0$, we find that 
\begin{eqnarray}
    0 &=& F(r)^{\frac{-2}{n-3}} G(r)^{\frac{n-4}{n-3}}  \Bigg( \frac{n-2}{8(n - 3)} \bigg( - 4\frac{F'^2}{F^2}  + \frac{G'^2}{G^2}  -  \frac{4(n - 3)^2 }{G}   \bigg)  +  \frac{1}{4} \frac{C^2}{G^2}   \Bigg)   \nonumber \\0 &=&  \frac{F'^2}{F^2}  - \frac{1}{4} \frac{G'^2}{G^2}  +  \frac{(n - 3)^2 }{G}   -  \frac{n - 3}{2(n - 2)} \frac{C^2}{G^2}
\end{eqnarray} We completed the derivation of equations \eqref{eom1} and \eqref{eom2}. \\
Now we derive the Ellis-Bronnikov class solution of this metric, in the case that $G(r)$ has no real roots, its solution for \eqref{master} can be written as 
\begin{equation}
    G(r) = \big( n - 3 \big)^2 r^2 + G_0
\end{equation} then
$$G' = 2(n-3)^2 r$$ equation \eqref{eomA} becomes 
\begin{equation}
    \frac{F''}{F} - \frac{F'^2}{F^2}  + \frac{2(n-3)^2 r}{(n - 3)^2 r^2 + G_0} \frac{F'}{F} = 0
\end{equation}

The solution of $F(r)$ is given by,
\begin{equation}
    F(r) = F_0  \exp  \Bigg( \beta \arctan{\bigg( \frac{r}{R} \bigg) } \Bigg)
\end{equation} where we defined the constant $G_0$ as 
\begin{equation}
    G_0 = \big( n-3 \big)^2 R^2
\end{equation} $F_0$ and $\beta$ are integration constants, in particular, we absorbed a constant $\frac{1}{R}$ in front of the $\arctan$ into the constant $\beta$.\\
Now going back to equation \eqref{eom2}, we find the constant $C$ to be
\begin{equation}
    C^2 = 2R^2 \big( n-2 \big) \big(n-3 \big)^3 \Big( 1+\beta^2 \Big)
\end{equation}
Use equation \eqref{phi}, $\phi' = \frac{C}{G}$, we find the axion field to be 
\begin{equation}
    \phi (r) = \phi_0 \pm \sqrt{\frac{2( n - 2 ) \big( 1+\beta^2 \big)}{n-3}} \arctan{\bigg( \frac{r}{R} \bigg)}
\end{equation} Then we make the coordinate change to put the metric into a better form. For convenience, let's simply relabel the original coordinates $t$ and $r$ as $\bar{t}$ and $\bar{r}$, then
\begin{eqnarray}
    \dd s^2 &=& -\frac{1}{F(\bar{r})^2} \dd \bar{t}^2 + F(\bar{r})^{\frac{2}{n-3}}G(\bar{r})^{-\frac{n-4}{n-3}}  \bigg(  \dd \bar{r}^2 + G(\bar{r}) \dd \Omega^2 \bigg)   \nonumber \\
    F(\bar{r}) &=& F_0 \exp \bigg( \beta \arctan{\Big( \frac{\bar{r}}{R} \Big) } \bigg)  \\
    G(\bar{r}) &=& \big( n-3 \big)^2 \Big( \bar{r}^2 + R^2 \Big)   \nonumber 
\end{eqnarray} We can write the metric explicitly as
\begin{eqnarray}
    \dd s^2 &=& - e^{-2 \beta \arctan{ ( \bar{r}/R ) } } \frac{\dd \bar{t}^2}{F_0^2}    \\
            &+& e^{ 2 \beta \arctan{ ( \bar{r}/R )} / (n-3) } \big( ( n-3 )^2 ( \bar{r}^2 + R^2 ) \big)^{-(n-4)/(n-3)}  \bigg(  \dd \bar{r}^2 +  ( n-3 )^2 \big( \bar{r}^2 + R^2  \big) \dd \Omega^2 \bigg)   \nonumber 
\end{eqnarray}
Now we express $\bar{t}$ and $\bar{r}$ in terms of the new coordinates $t$ and $r$, 
\begin{equation}
    \bar{t} = F_0 t,~~ \bar{r} = \frac{2R r^{n-3}}{M},~~ M = 2(n-3) R F_0
\end{equation}
Defining
\begin{eqnarray}
    U(r) &:=& \arctan{\bigg( \frac{2 r^{n-3}}{M} \bigg)} \\
    V(r) &:=& 1 + \frac{M^2}{4 r^{2(n - 3)}}
\end{eqnarray} we can write the axion fields as
\begin{equation}
    \phi(r) = \phi_0 \pm \sqrt{ \frac{2(n-2)\big( 1+ \beta^2 \big)}{n-3}} U(r)
\end{equation}
Also we have 
\begin{eqnarray}
    V(r)^{1/(n-3)} r^2 &=& \qty( \qty( 1 + \frac{M^2}{4r^{2(n-3)}} ) r^{2(n-3)} )^{1/(n-3)}  \nonumber \\
    &=& \qty\Big( r^{2(n-3)} + (n-3)^2 R^2  )^{1/(n-3)} \nonumber \\&=& G(\bar{r})^{-(n-4)/(n-3)} G(\bar{r})
\end{eqnarray} and
\begin{eqnarray}
     \qty\Big( (n-3)^2 \qty\big(\bar{r}^2 + R^2 ) )^{-\frac{n-4}{n-3}} \dd \bar{r}^2 &=& \qty\Big( (n-3)^2 R^2 + r^{2(n-3)} )^{-\frac{n-4}{n-3}}  \frac{4(n-3)^2 R^2}{M^2} r^{2(n-4)} \dd r^2 \nonumber \\
     &=& \qty( \frac{(n-3)^2 R^2 }{r^{2(n-3)}} + 1 )^{-\frac{n-4}{n-3}}  \dd r^2   \nonumber  \\
     &=& V(r)^{1/(n-3)} \frac{\dd r^2}{V(r)}
\end{eqnarray}
The metric is now written as 
\begin{equation}
    \dd s^2 = -e^{-2\beta U(r)} \dd t^2 + e^{2\beta U(r)/(n-3)} V(r)^{1/(n-3)} \bigg( \frac{\dd r^2 }{V(r)} + r^2 \dd \Omega^2 \bigg) 
\end{equation}

\end{document}